\documentclass[aps,prb,preprint,amsmath,amssymb,showpacs,prl,unsortedaddress]{revtex4-1} 
\usepackage{epsf}      
\usepackage{graphicx}
\usepackage{gensymb}
\usepackage{sidecap}

\newcommand{\etal}{{\it et al.}}

\begin{document}

\title{Pairing, Pseudogap and Fermi Arcs in Cuprates}

\author{Adam Kaminski}
\affiliation{Division of Materials Science and Engineering, The Ames Laboratory, U.S. DOE}
\affiliation{Department of Physics and Astronomy, Iowa State University, Ames, Iowa 50011, USA}
\author{Takeshi Kondo}
\affiliation{Institute for Solid State Physics, The University of Tokyo, Kashiwa, Chiba Japan}
\author{Tsunehiro Takeuchi}
\affiliation{Department of Crystalline Materials Science, Nagoya University, Nagoya 464-8603, Japan}
\author{Genda Gu}
\affiliation{Condensed Matter Physics and Materials Science Department, Brookhaven National Laboratory, Upton, New York 11973, USA}

\date{\today}     

\begin{abstract}
We use Angle Resolved Photoemission Spectroscopy (ARPES) to study the relationship between the pseudogap, pairing and Fermi arcs in cuprates. High quality data measured over a wide range of dopings reveals a consistent picture of Fermiology and pairing in these materials. The pseudogap is due to an ordered state that competes with superconductivity rather then preformed pairs.  Pairing does occur below T$_{pair}\sim150K$ and significantly above Tc, but well below T* and the doping dependence of this temperature scale is distinct from that of the pseudogap. The d-wave gap is  present below T$_{pair}$, and its interplay with strong scattering creates ``artificial'' Fermi arcs for  T$_c\le$ T $\le$ T$_{pair}$. However, above T$_{pair}$, the pseudogap exists only at the antipodal region. This leads to presence of real, gapless Fermi arcs close to the node. The length of these arcs remains constant up  to T*, where the full Fermi surface is recovered. We demonstrate that these findings resolve a number of seemingly contradictory scenarios.
\end{abstract}

\maketitle

\section{Introduction}

The pseudogap is one of the most interesting aspects of the physics of the cuprates, second only to their unusually high transition temperatures. Since its discovery \cite{Warren, Takigawa, Homes, Timusk}, the origin of the pseudogap and its relationship to high temperature superconductivity has been a subject to intense debate over the last two decades\cite{MikeFoe}. One of the most popular models is based on the existence of pre-formed pairs \cite{Emery}, but other models invoking the presence of an ordered state have been proposed too\cite{Hsu, Varma97, Varma99, Chakravarty}. A circular dichroism in the pseudogap state was discovered by early ARPES experiments \cite{Kaminski CD} and presence of staggered magnetic field was detected using high precision magneto-optical Kerr effect \cite{Kapitulnik} as well as neutron scattering measurements\cite{Burge}. Early STM measurements revealed presence of checker board pattern consistent with charge ordering in the cores of the vortexes \cite{Hoffman} then in heavily under doped samples\cite{Hanaguri} and moderately doped samples\cite{EricCDW}. More recently series of high precision X-ray scattering experiments detected fluctuating charge density order within pseudo gap state\cite{Ghiringhelli,LeTacon,daSilvaNeto}. All these findings point to the fact that pseudo gap is a manifestation of an ordered states or perhaps interplay of two ordered states involving charge and magnetic moment.

Angle Resolved Photoemission Spectroscopy (ARPES) has played a very important role in elucidating the electronic properties of the cuprates \cite{DamascelliReview, CampuzanoReview}. Examples of ARPES data measured on Bi2201 samples along with a diagram of the Fermi surface are shown in Fig. 1. Indeed, due to it's unique momentum resolution, ARPES was one of the techniques of choice in studies of the pseudogap \cite{HongPseudogap, LoeserPseudogap} and the one that led to discovery of Fermi arcs \cite{NormanNature}. The classical definition of the Fermi surface is a set of closed contours in momentum space that separates the occupied and unoccupied states. Fermi arcs, being disconnected and possessing by definition ``end points'' are a highly unusual concept in "classical" condensed matter physics. 

Many of the early studies of the pseudo gap and Fermi arcs relied on line shape analysis, primarily Energy Distribution Curves (EDCs). The signature of the pseudogap in 
symmetrized EDCs is shown in Fig. 2. The spectra is considered as ``gapped'', when a dip or flat top are present  at E$_f$. A single peak  at E$_f$ is considered evidence for no gap. This conventional approach led to several important  advances such as measurements of the superconducting gap anisotropy. Yet, at the same time, it has limitations, particularly when broad spectral peaks are present in the pseudogap state at the antinode. Here we revisit these issues using very high quality data and a novel quantitative analysis. This approach is more objective, has higher sensitivity to spectral changes and is rooted in the basic definition of the energy gap, based on the density of states.  

We demonstrate that at low temperature in optimally and under doped cuprates, two distinct gaps are present: the superconducting gap and a pseudogap. They represent different energy scales that evolve in distinct ways with momentum, temperature and doping. Furthermore, we show that the pseudogap competes with superconductivity by depleting the low energy electrons in the antinodal region, that otherwise would form pairs below T$_c$. In the under doped materials therefore, only a small portion of the Fermi surface close to the nodal region contributes to the superfluid density. Our quantitative approach provides evidence of pairing above T$_c$ and allows us to detect the onset temperature of pairing T$_{pair}$ that is significantly lower than the pseudo gap temperature T*. This  naturally explains earlier conclusions about the extension of the d-wave gap above T$_c$. Our data strongly supports theoretical proposals that Fermi arcs below T$_{pair}$ are an artifact due to the interplay between a d-wave gap and strong scattering effects \cite{ChubukovScattering}. Using this quantitive approach we are also able to demonstrate that above T$_{pair}$, real, gapless Fermi arcs exist in the nodal region and are likely due to an ordered state\cite{KAMINSKITRSB, MagneticOrder,  EricCDW, EricFS, Hanaguri} responsible for the pseudogap.

\section{Experiments}

Optimally doped Bi$_2$Sr$_2$CaCu$_2$O$_{8+\delta}$ (Bi2212) single crystals with $T_{\rm c}$=93K (OP93K) and  (Bi,Pb)$_2$(Sr,La)$_2$CuO$_{6+\delta}$ (Bi2201) single crystals with $T_{\rm c}$=32K (OP32K) were grown by the conventional floating-zone (FZ) technique. We partially substituted Pb for Bi in Bi2201 to suppress the modulation in the BiO plane, and avoid contamination of the ARPES signal with diffraction images due to the superlattice: the outgoing photoelectrons are diffracted at the modulated BiO layer, creating multiple images of the bands and Fermi surface that are shifted in momentum. The modulation-free samples enable us to precisely analyze the ARPES spectra. 

ARPES data was acquired using a laboratory-based system consisting of a Scienta SES2002 electron analyzer and GammaData helium UV lamp. All data were acquired using the HeI line with a photon energy of 21.2 eV. The angular resolution was $0.13^\circ$ and $\sim 0.5^\circ$ along and perpendicular to the direction of the analyzer slits, respectively.  The energy corresponding to the chemical potential was determined from the Fermi edge of a polycrystalline Au reference in electrical contact with the sample. The energy resolution was set at $\sim10$meV - confirmed by measuring the energy width between 90$\%$ and 10 $\%$ of the Fermi edge from the same Au reference. 
\setcounter{figure}{0}
\begin{figure}
\centerline{\includegraphics[width=5in]{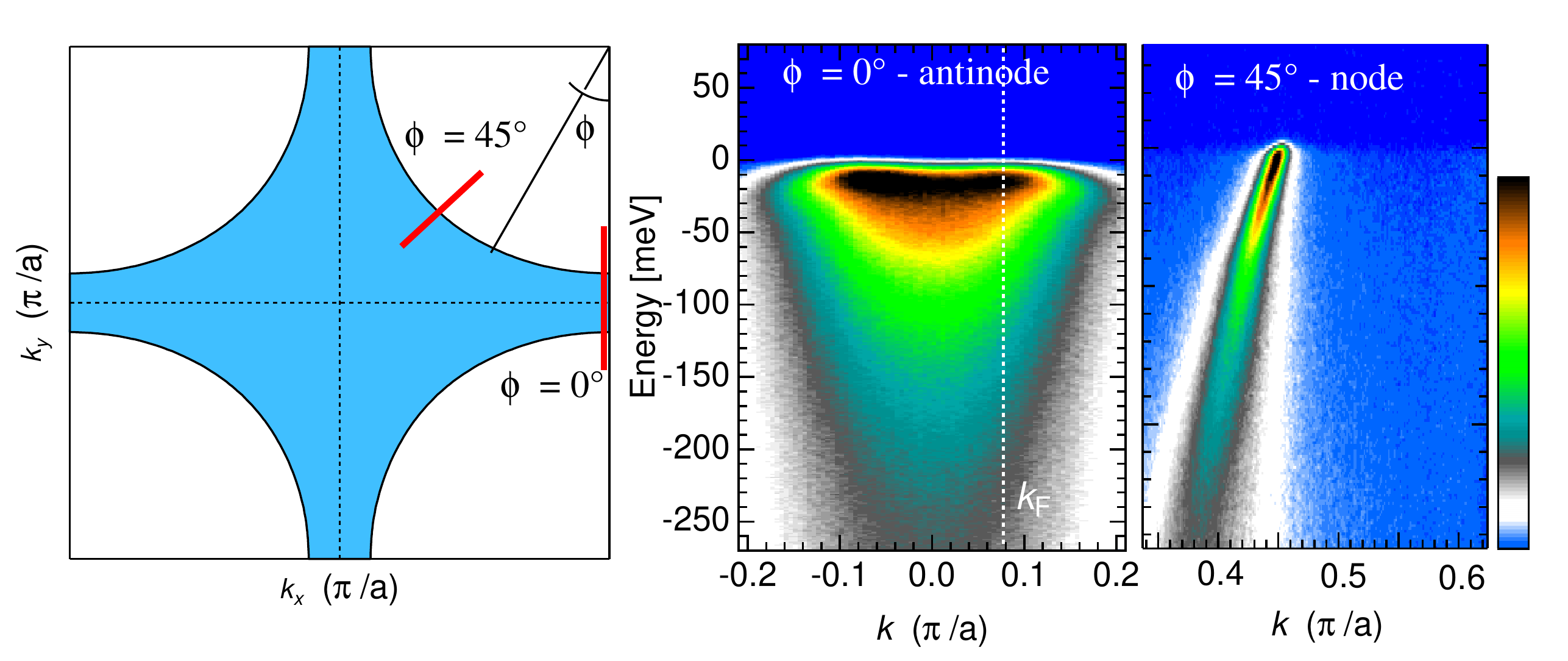}}
\caption{a) Plot of the Brillouin zone and Fermi surface contour for Bi2201 showing the definition of the Fermi surface angle. b) ARPES intensity plot along the antinodal cut (marked with a red line for $\phi=0$. c) ARPES data along the nodal direction ($\phi=45$). The data was measured at T=11K.}
\label{landfig}
\end{figure}

Custom designed refocusing optics enabled us to accumulate high statistics spectra in a short period of time with no sample surface aging from the absorption or loss of oxygen. Special care was taken to purify the helium gas supply for the UV source to remove even the smallest trace of contaminants that could contribute to surface contamination. Typically no changes in the spectral lineshape of samples were observed in consecutive measurements performed over several days. We constructed a sample manipulator with the tilt and azimuth motions mounted on a two stage closed cycle He refrigerator. To measure the partial density of states along a direction perpendicular to the FS we controlled the sample orientation in-situ. Specially designed temperature management system allowed us to rapidly change sample temperature. This was critical for obtaining high density of data points in temperature scan and further limit sample aging. Measurements were performed on several samples and we confirmed that all yielded consistent results. 

\section{Results and discussion}

\subsection{Two energy gaps}

\begin{figure}
\centerline{\includegraphics[width=5.5in]{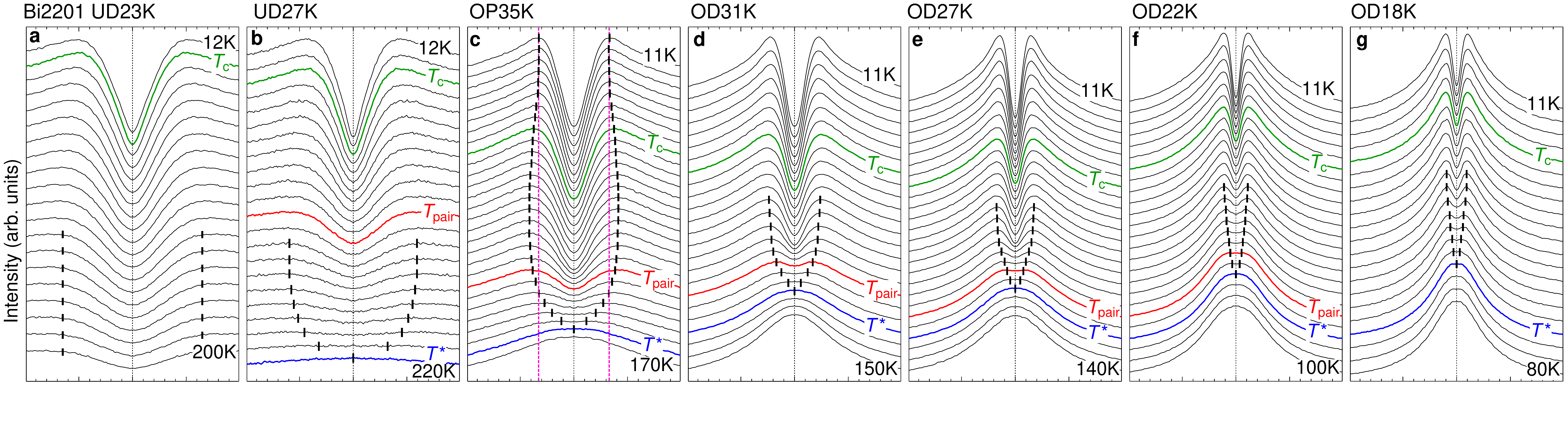}}
\caption{Signature of the pseudogap in ARPES data at the antinode. The curves represent symmetrized EDCs measured at different temperatures and doping levels (panels). A dip at E$_f$ and double peak structure indicates the presence of an energy gap. This gap persists well above T$_c$ and the gap closing temperature T* increases with decreasing doping.}
\label{landfig}
\end{figure}

The issue of one vs two energy gaps is directly tied to the origin of the pseudogap. If the pseudogap is a state of pre-formed pairs, then naturally the energy gap arises from pairing and has the same form for all temperatures below T* as reported in number of studies \cite{Nakayama, Chatterjee, VallaScience, Johnson}. However several other studies including  Raman Spectroscopy, ARPES and STM  \cite{Raman,ShenScience,ShenNature,Twogap,EricTwoGap} reveal certain differences in the behavior of the energy gap close to the node compared to the antipodal region. More precisely, the spectral gap follows the behavior expected of d-wave pairing gap only in regions of momentum space close to the node. In the antinodal regions there are significant deviations from d-wave behavior. For example, the antipodal gap magnitude continuously increases with decreased doping, despite a decrease of T$_c$ and it is almost temperature independent up to T*. We illustrate this in Fig. 3, where the magnitude of the ``spectral gap'' (we use that term to refer to the gap extracted from the data to avoid bias) is plotted for three doping levels, below and above T$_c$. The dotted lines mark the expected behavior of a d-wave order parameter. In the overdoped sample (panel c), at low temperatures, the spectral gap follows exactly the predictions based on a d-wave order parameter. Above T$_c$, the gap has a similar magnitude in the antipodal region, but vanishes near the node, unlike a d-wave gap. At optimal doping, the low temperature data close to the node, fits d-wave very well, but deviates  strongly in the antinodal region. This becomes more extreme in underdoped samples (panel a). It is also worth noting that the maximum value of the spectral gap at the antinode for the UD23K sample is larger by a factor of 2 than that of the OP35K sample, despite an actual decrease of the superconducting critical temperature. These data suggest that the spectral gap in under doped cuprates has two components, namely a d-wave superconducting gap and a pseudogap. The two gaps have different momentum, doping and temperature dependences. The superconducting gap follows the doping dependence of T$_c$, vanishes well below T* and has pure d-wave momentum dependence. The pseudogap, in contrast, continuously increases with decreased doping, persists up to T* and only exists close to the antinode. 
\begin{figure}
\centerline{\includegraphics[width=5in]{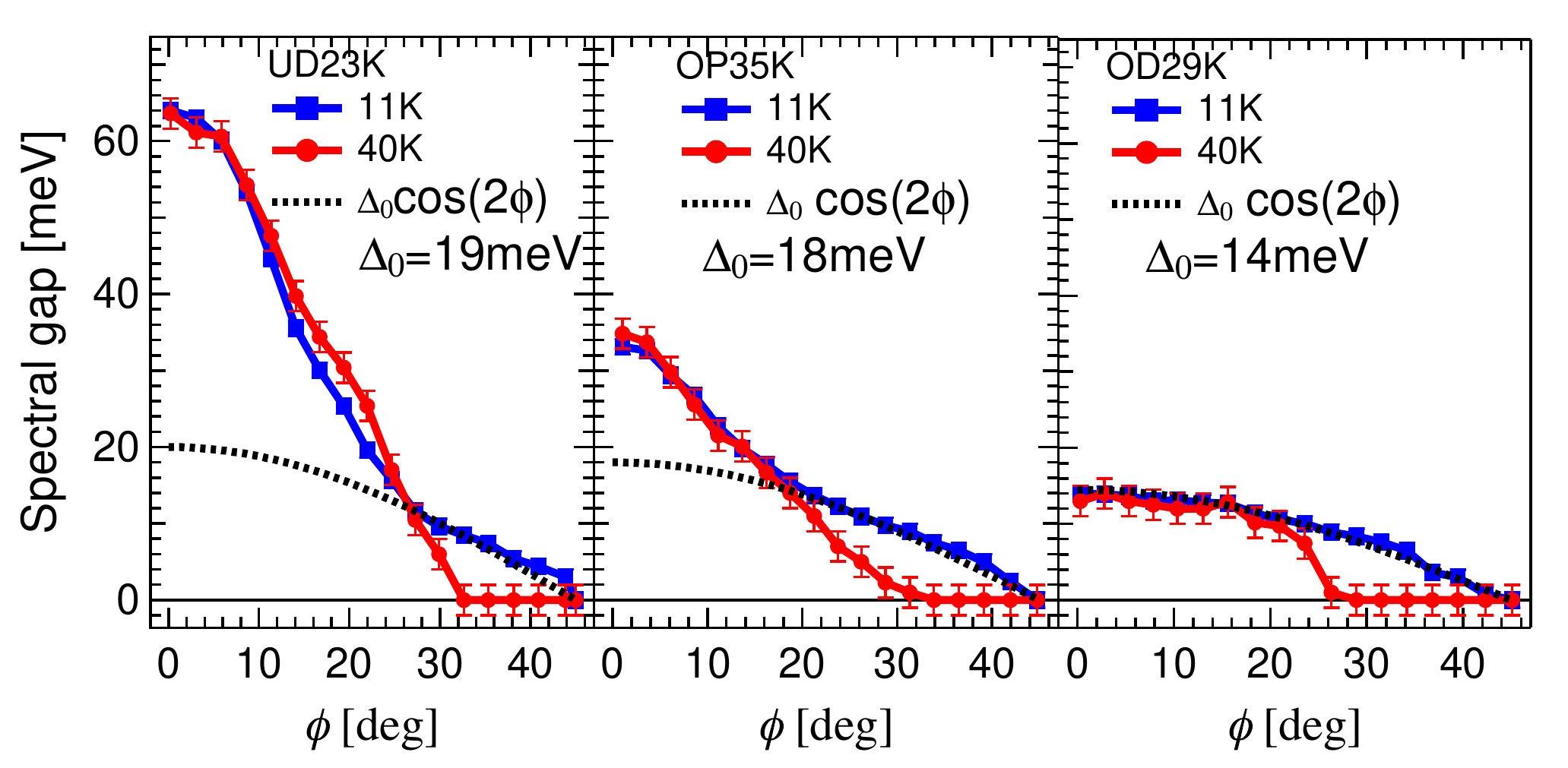}}
\caption{Momentum dependence of the spectral gap for three doping levels measured below and above T$_c$. The dotted line is expectation of d-wave order parameter.}
\label{landfig}
\end{figure}

Perhaps the most convincing evidence for the existence of two distinct gaps in the cuprates is found by examining very carefully the temperature dependence of very high quality EDC data (Fig. 4). Blue lines mark the peak positions and therefore the magnitude of the gap. As the temperature is increased, the gap actually increases in magnitude, only to vanish close to the T*. This behavior is illustrated in panel b, where we plot the magnitude of the spectral gap as a function of temperature for several doping levels. Such a violation of monotonicity is completely inconsistent with a single gap picture and can be only explained by the two gap scenario. More precisely, below T$_c$, sharp quasiparticle peaks mark the location of the d-wave pairing gap. As T$_c$ is approached, those peaks vanish and reveal a second gap of larger magnitude that persists up to T* and it is this therefore that is the proper pseudogap. In heavily underdoped samples (e. g. UD25K) the pseudogap is so large, that no quasiparticle peaks are present at the antinode even at the lowest temperature. In overdoped samples, the pseudogap is much smaller than the pairing gap, which dominates the spectra and preserves the monotonicity. In the range of doping where the magnitude of the pseudogap is larger than pairing gap, but not large enough to wipe out the quasiparticle peaks, non-monotonic behavior is observed. 

\begin{figure}
\centerline{\includegraphics[width=5in]{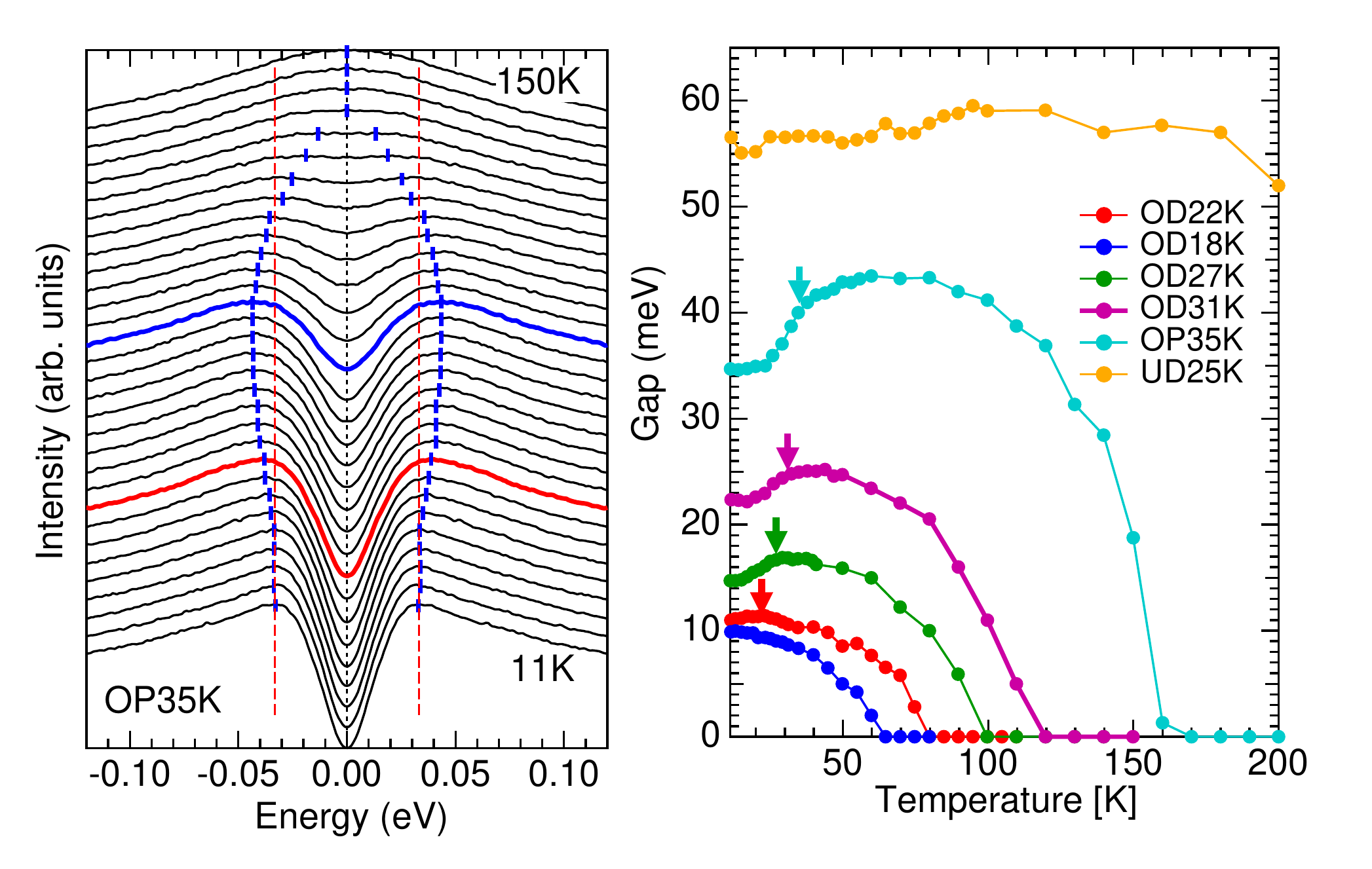}}
\caption{Detailed temperature dependence of the spectral gap. a) symmetrized EDCs as a function of temperature at the antinodal crossing in optimally doped sample. The value of the gap extracted from the data is marked by blue lines. Value of the spectral gap as a function of temperature for several doping levels. Note that in data close to optimal doping, the gap first increases upon warming, then decreases to zero at temperatures much higher than T$_c$.}
\label{landfig}
\end{figure}

\subsection{Competition between the pseudogap and superconductivity}
\begin{figure}
\centerline{\includegraphics[width=5in]{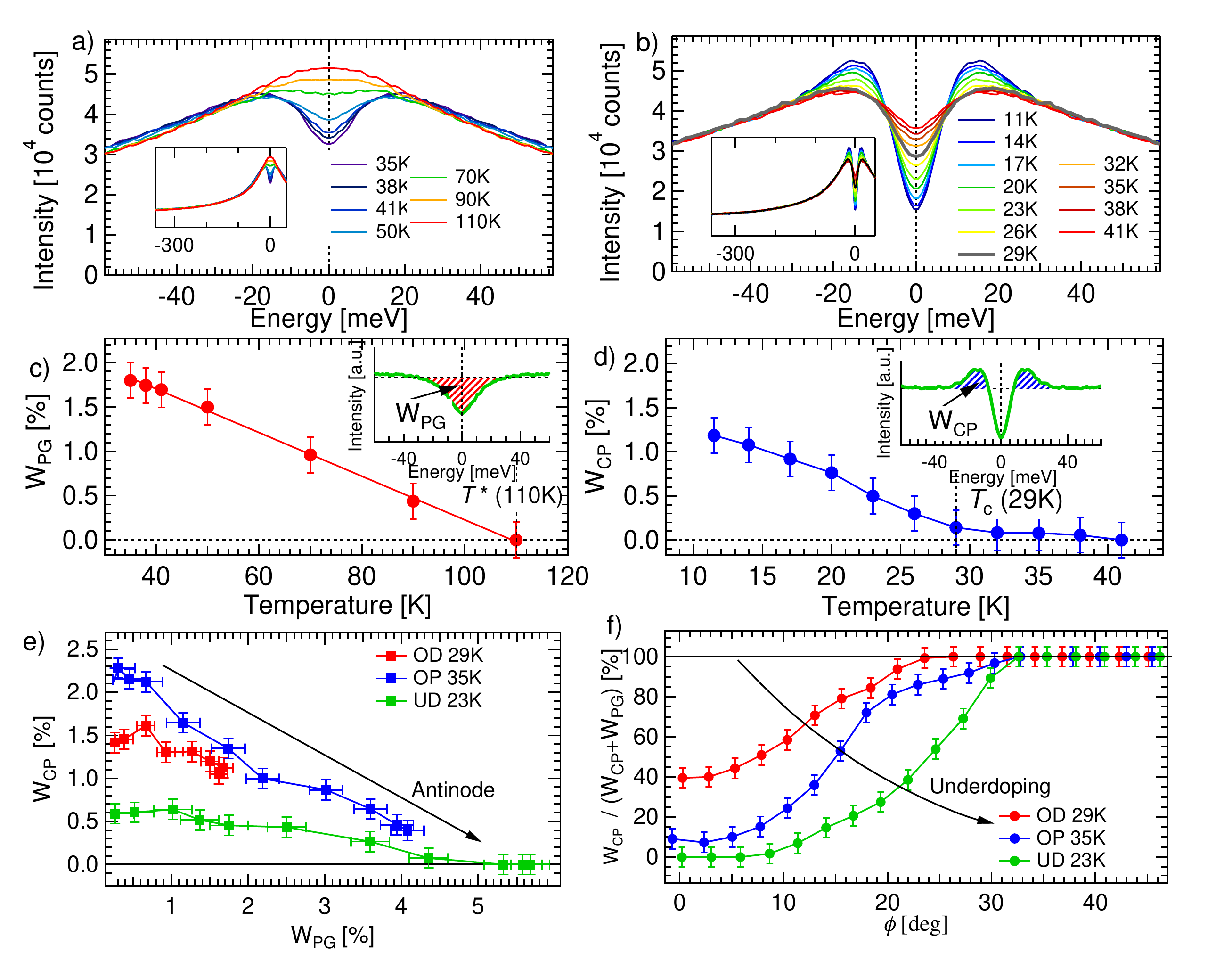}}
\caption{Evolution of symmetrized EDC data in pseudo gap state b) and superconducting state c). The spectral weight lost due to pseudo gap state is defined in the insert and its temperature dependence is shown in panel c). The superconducting quasiparticle weight is defined in the insert and its temperature dependence is shown in panel d). e) Superconducting quasiparticle weight plotted against weight lost due to the pseudogap for each of the momentum points along the Fermi surface. f) Fraction of the quasiparticle weight to the total change in the spectral weight as a function of momentum for three dopings. }
\label{landfig}
\end{figure}

Now that we have demonstrated the existence of two energy scales in the cuprates, a natural question arises about their mutual relationship. The most direct way to answer this is to look how these two energy scales affect the spectral properties. As we mentioned in the introduction, conventional line shape analysis is difficult to utilize here. Instead, we use the quantitative analysis shown in Fig. 5ab \cite{KondoCompetition}. In the pseudogap state, the spectral weight near E$_f$ decreases upon cooling. The change in this spectral weight (marked in red in the insert of panel 5c) is a good measure of the ``strength'' of the psudogap. In the superconducting state, however, the  formation of quasiparticle peaks occurs at the pairing gap edge (marked in blue areas in panel 5d). These peaks signify the contribution to a superfluid density from this region of momentum space\cite{FengScience}. The temperature dependence of both quantities  is shown in panels 5 c and d. The pseudogap weight decreases linearly to zero at T*, while the quasiparticle weight vanishes at T$_c$. To study the relationship between the two quantities we plot in Fig. 5e the weight of the quasiparticle peaks versus pseudogap weight for momentum points, where they are non-zero.  For all three dopings, there is a strong anti-correlation between the two quantities. The superconducting quasiparticle peak weight decreases as the pseudogap weight increases at all momentum points and all three dopings. Such a strong anti-correlation is definitive proof that the two phenomena compete for available states near E$_f$. In Fig. 5f we look at the momentum dependence of the normalized weight of the quasiparticle peaks. If the pseudogap really was a state of pre-formed pairs, the weight transferred from near E$_f$, should re-condense into coherent pairs below T$_c$, thus the quantity shown in Fig. 5f should remain at 100\% for all momentum points. Instead we observe a strong depletion of the quasiparticle weight in the antipodal region (where the pseudogap is strongest), that increases with decreased doping. In heavily underoped sample (green curve), the pseudogap completely dominates the antipodal region and there are no quasiparticle peaks, thus only areas close to the node participate in superconductivity. This again signifies that the pseudogap competes with the formation of coherent pairs. These conclusions are in good agreement with thermodynamics studies\cite{Rustem} and were recently confirmed by another ARPES study \cite{Vishik}.

\subsection{Pairing above T$_c$}

Competition of the pseudogap with superconductivity does not preclude the formation of Cooper pairs above T$_c$. Indeed, several thermodynamical and transport studies report signatures of pairing above T$_c$, which do not however persist up to T* \cite{OngPRB, OngDiamagnetism, Terahertz, Josephson, Panagopoulos, Tallon}. We find spectroscopic evidence of pair formation and were able to establish the value of the pairing temperature using our quantitative approach\cite{KondoPairing}. In Fig. 6 we plot a detailed temperature dependence of the spectral weight in Bi2201 at E$_f$ (integrated within 10 meV) and antinodal crossing for several doping levels. In the heavily underdoped sample (panel a), where the spectrum is completely dominated by the pseudo gap (as demonstrated in the section above), this weight is linear with temperature. Such linear behavior is a signature of the pseudogap. As we increase the doping, there is a deviation from such linear form, and the weight at E$_f$ decreases faster below certain temperature (marked by the red arrow and referred to as T$_{pair}$). We note that there are no other striking features in this dependence as we cross T$_c$. This is a strong evidence that pairing occurs already at T$_{pair}$ and partially coexists with the pseudogap. With increased doping, the weight lost to the pseudogap (blue area) decreases and the area related to pairing increases - signifying an increase in superfluid density. This is consistent with direct measurements of the superfluid density by $\mu$SR \cite{Rustem}. We also note that the pairing temperature reaches 150K in Bi2212, which likely sets the maximum achievable critical temperature in cuprates. This also resolves the controversy between the existence of two distinct energy scales and reports of a continuous evolution of the gap across T$_c$ \cite{Nakayama, Chatterjee, VallaScience, Johnson}. Below T$_{pair}$, a d-wave pairing gap is present in the spectra, therefore no dramatic changes in the gap properties are observed. Only at  higher temperatures, do distinct features of the pseudogap manifest themselves.

\subsection{Real Fermi arcs}
\begin{figure}
\centerline{\includegraphics[width=5.5in]{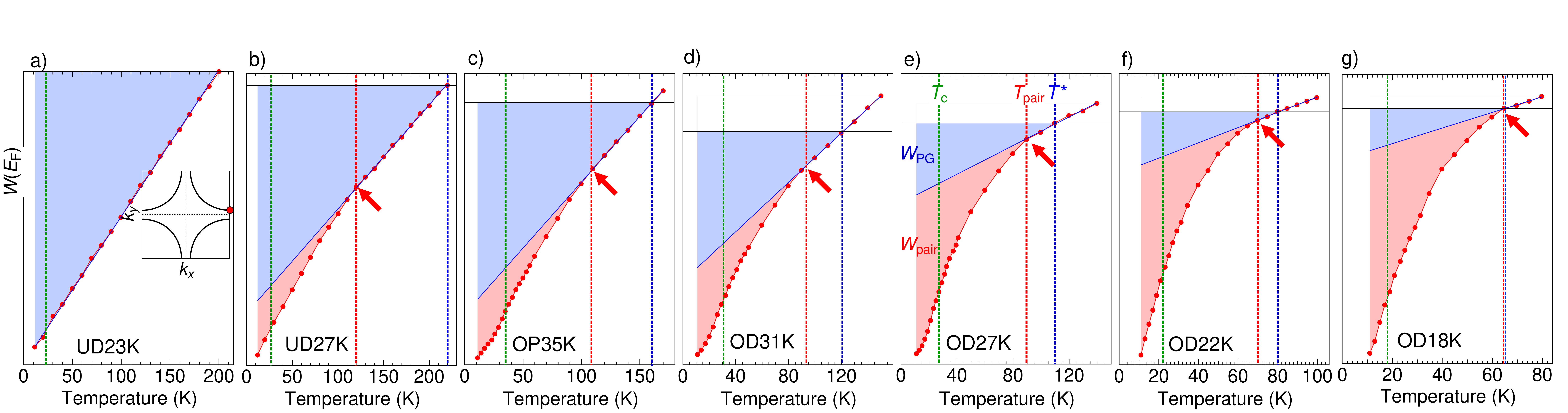}}
\caption{Temperature dependence of the spectral weight at E$_f$ for different doping levels. The blue line is an extrapolation of the linear behavior present at high temperatures. The arrow marks the departure from linear behavior. The blue area is enclosed by the linear extrapolation. Additional weight lost (beyond linear behavior - marked pink) is due to pairing.}
\label{landfig}
\end{figure}

The discovery of spectroscopic signatures of pairing is very closely related to the issue of Fermi arcs. Earlier studies reported a linear dependence of the length of these arcs with temperature \cite{KanigelNature}. This was interpreted in a quite natural way as the interplay of a d-wave pairing gap and strong scattering that exists above T$_c$ \cite{ChubukovScattering}. When the scattering rate exceeds the magnitude of the gap, symmetrized spectra lack the dip at E$_f$ and appear gapless. Since the d-wave gap is smallest in the nodal region, this part of the Fermi surface appears gapless, while spectra away from the node still bear the signature of the gap. As the temperature increases, so does the scattering rate and a larger portion of the Fermi surface appears "normal". This gives the appearance of arcs expanding with increasing temperature. \cite{ChubukovScattering}. This scenario was recently confirmed by follow-up ARPES experiments\cite{Reber}, which demonstrated that a d-wave gap is still present at all Fermi momentum points up to  150K (which is of the order of our T$_{pair}$). An interesting question arises. What is the Fermi surface at higher temperatures in the absence of pairing, where the pseudo gap is still present? 
\begin{figure}
\centerline{\includegraphics[width=5.5in]{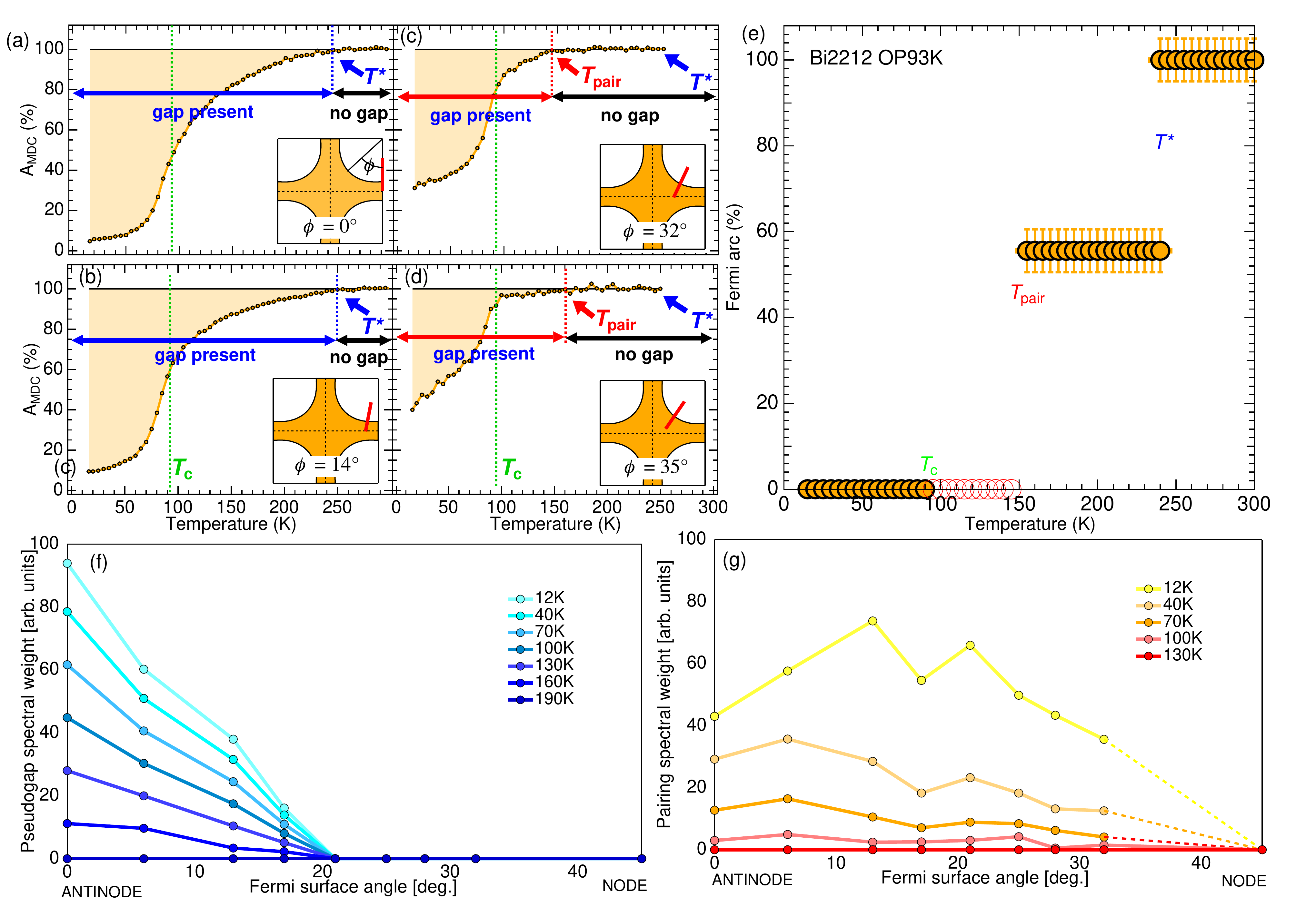}}
\caption{Measurements of the gap opening temperature using a DOS approach. panels (a-d) show the temperature dependence of the MDC weight. When this quantity begins to decrease it implies the opening of the gap. In the antinode region (panels a, c) this occurs at T*, while in the nodal region, the gap opens at a much lower temperature T$_{pair}$, however still above T$_c$. Panel e) shows the length of the real Fermi arc as a function of temperature determined using this method. Open red circles are plotted based on conclusions from Ref. \cite{Reber}}
\label{landfig}
\end{figure}

To examine this issue in detail we again utilize our quantitative approach \cite{KondoArcs}. The most sensitive probe of the opening of an energy gap is the density of states (DOS) at E$_f$. To obtain momentum dependent information we note that the area of the MDC at E$_f$ along a cut in momentum space represents a contribution to the DOS(E$_f$). When this quantity is integrated over the Brillouin zone, it is directly proportional to DOS(E$_f$) modulo matrix elements. To detect the opening of an energy gap we monitor the area of the MDC (A$_{MDC}$) along a cut as a function of temperature. In the absence of gap, this quantity should be constant, since it is independent of the scattering rate (only the width of the MDC increases with the scattering rate). As the gap opens, this quantity should decrease, and any departure from a constant value marks the gap opening temperature. We use this method to detect the gap opening temperature along the Fermi surface of Bi2201 and Bi2212. Examples of such plots are shown in Fig. 7 a-d. Close to the antinode, the pseudo gap opens at T*. A$_{MDC}$ is constant above T* ($\sim$ 250K signifying the absence of a gap and it decreases below this temperature as the pseudogap opens. This quantity decreases faster once T$_{pair}$ is reached and is consistent with the conclusions drawn in the previous section. Close to the node, A$_{MDC}$ remains constant well below T* and it only starts to decrease once T$_{pair}$ is reached. This occurs at all momentum points for $\phi\ge20 ^\circ$. In contrast, the pseudogap opens over the remainder of the Fermi surface exactly at T*. This proves that real, gapless Fermi arcs exist  above T$_{pair}$. Furthermore, since the pairing gap opens at all points in this part of the Fermi surface below T$_{pair}$, the length of this arc remains constant between T* and T$_{pair}$. Indeed in Fig. 7e we plot the length of the Fermi arc, extracted by examining the gap opening temperature at 8 points along the Fermi surface. As the temperature is lowered below  T*, the full metallic Fermi surface collapses to gapless Fermi arcs. The length of these arcs remains constant down to T$_{pair}$, below which the arcs collapse to the node of a d-wave gap. The density of our data points close to the node allow us to put an upper limit on the size of the arc below T$_{pair}$, but  we know from Ref. \cite{Reber}, that the Fermi surface below 150K has a full d-wave gap. To demonstrate the universality of our findings, we plot in Fig. 7f, the weight lost to the pseudogap as a function of momentum and temperature (in a similar fashion to Fig. 6, but extracted from MDC areas).  All data points extrapolate to $\phi\sim20 ^\circ$, demonstrating that this endpoint of the arc is independent of  temperature. Similarly in Fig. 7g, we plot the area representing paired states. Again the points extrapolate to the nodal point for all temperatures below T$_{pair}$, demonstrating that a d-wave gap is present with a single gapless node per quadrant.
\begin{figure}
\centerline{\includegraphics[width=5in]{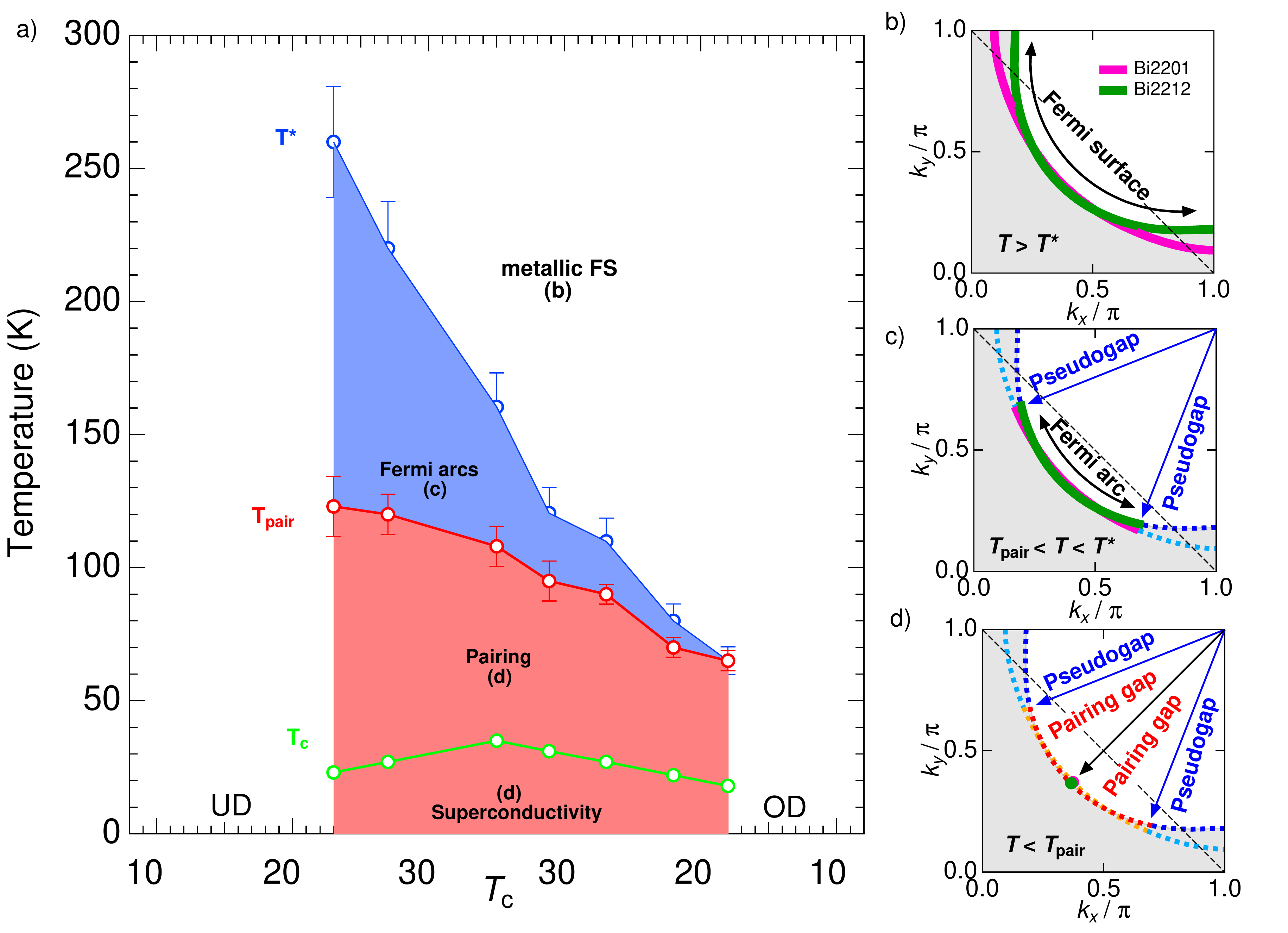}}
\caption{Phase diagram of Bi2201 with superconducting, pairing and pseudo gap temperatures. In the blue area only the pseudo gap is present in the samples and real, gapless Fermi are exist. In the red area, a pseudogap coexists with pairing and a d-wave order parameter gaps Fermi surface. Panels (b-d) schematically show gapless Fermi surface in three key areas.}
\label{landfig}
\end{figure}

\section{Conclusions}

In summary, we arrived at a consistent Fermiology of pseudogap state in cuprates by utilizing a quantitative approach to ARPES data. This picture reconciles most major controversies and is summarized in Fig. 8. In panel A we plot the partial phase diagram with measured superconducting, pairing and pseudogap temperatures. The Fermi surface in key regions are plotted schematically in panels (b-d). We demonstrated that the pseudogap and pairing gap are two distinct energy scales that compete for low energy states. Pairing in the cuprates occurs well above T$_c$, but also significantly below T*. The pairing temperature T$_{pair}\sim150K$ likely sets maximum achievable superconducting transition temperature in the cuprates. Below T$_{pair}$ a d-wave gap is present at all points of the Fermi surface and gapless Fermi arcs are absent. Above  T$_{pair}$, the pseudogap dominates the spectral properties and leads to the formation of real, gapless Fermi arcs in the nodal region. The length of these arcs do not change with temperature and their tips define a set of vectors that are likely the key to understanding the order responsible for the origin of the pseudogap.

\label{lastpage}

\end{document}